\documentclass[aps,prb,preprint,superscriptaddress]{revtex4}

\usepackage{graphicx}

\begin{document}


\title{Bandgaps and electronic structure of alkaline earth and post-transition metal oxides}

\author{J.~A.~McLeod}
\author{R.~G.~Wilks}
\affiliation{Department of Physics and Engineering Physics, University of Saskatchewan, 116 Science Place, Saskatoon, Saskatchewan S7N 5E2, Canada}
\email[Contact Author:]{john.mcleod@usask.ca}
\author{N.~A.~Skorikov}
\author{L.~D.~Finkelstein}
\affiliation{Institute of Metal Physics, Russian Academy of Sciences-Ural Division, 620219 Yekaterinburg, Russia}
\author{M.~Abu-Samak}
\affiliation{Department of Physics, Al-Hussein Bin Talal University, P.O. Box 20, Ma'an, Jordan}
\author{E.~Z.~Kurmaev}
\affiliation{Institute of Metal Physics, Russian Academy of Sciences-Ural Division, 620219 Yekaterinburg, Russia}
\author{A.~Moewes}
\affiliation{Department of Physics and Engineering Physics, University of Saskatchewan, 116 Science Place, Saskatoon, Saskatchewan S7N 5E2, Canada}

\date{\today}

\begin{abstract}
The electronic structure in alkaline earth \textit{Ae}O (\textit{Ae} = Be, Mg, Ca, Sr, Ba) and post-transition metal oxides \textit{Me}O (\textit{Me} = Zn, Cd, Hg) is probed with oxygen \textit{K}-edge X-ray absorption and emission spectroscopy. The experimental data is compared with density functional theory electronic structure calculations. We use our experimental spectra of the oxygen \textit{K}-edge to estimate the bandgaps of these materials, and compare our results to the range of values available in the literature. From the calculated partial DOS we conclude that the position of main O \textit{K}-edge X-ray emission feature in BeO, SrO and BaO is defined by the position of the \textit{np}-states of the cation while in the other oxides studied here the main O \textit{K}-edge X-ray emission feature is defined by the position of the \textit{(n-1)d} (for CaO) or \textit{nd}-states of the cation. 
\end{abstract}

\maketitle

\section{Introduction}
There have been many studies on the electronic structures of various metal oxides \cite{acke98, paganini02, chaudhary98, elfimov02, kenmochi04, kenmochi04_2, elfimov07, baltachea04}. Among these materials the alkaline earth oxides \textit{Ae}O (\textit{Ae} = Be, Mg, Ca, Sr, Ba) are considered to represent typical ionic crystals. These materials are insulators, and are important in a wide range of industrial applications from catalysis to microelectronics \cite{acke98, paganini02}. In particular, the catalytic properties of these materials are important in chemical engineering \cite{chaudhary98}. It was recently predicted that carbon and nitrogen doped MgO, CaO, and SrO exhibit ferromagnetism \cite{elfimov02, kenmochi04, kenmochi04_2, elfimov07}; this motivated our exploration of the electronic structure of these materials. Although there have been numerous bandstructure calculations performed for these compounds (see for example Reference \onlinecite{baltachea04} and references therein) experimental studies of the electronic structure of these compounds are scarce and focus mainly on photoemission measurements of the valence bands \cite{elfimov02, elfimov07, ochs96, cappellini01, sorokin76}.

The post transition metal oxides are II-VI type semiconductors. ZnO in particular is the subject of active research in optical and photo-electronic applications \cite{bagnall97, ozgur05, azzaz08}. CdO and HgO have similar electronic and opto-electronic applications \cite{liu03} and have found use in devices like solar cells \cite{champness95, sravani96}, transparent electrodes \cite{benko86}, and photo diodes \cite{kondo71}.

Since both alkaline earth oxides and post-transition metal oxides have similar valence configurations but exhibit different electronic properties, the combined study of the electronic structure and bandgap of both alkaline earth and post-transition metal oxides is of interest. By bandgap, we refer to the energy gap between the highest occupied state and the lowest unoccupied state, without any consideration of momentum difference, or interband excitonic states. These are important for the optical gap, which we do not investigate here.

In this manuscript we probe the occupied and unoccupied oxygen \textit{2p} density of states (DOS) of the alkaline earth oxides \textit{Ae}O (\textit{Ae} = Be, Mg, Ca, Sr, Ba), and post-transition metal oxides \textit{Me}O (\textit{Me} = Zn, Cd, Hg) using synchrotron-excited soft X-ray emission spectroscopy (XES) and X-ray absorption spectroscopy (XAS). The measurements are compared with \textit{ab initio} electronic structure calculations. 

The preferred method for estimating the bandgap is to employ a combination of X-ray photoemission spectroscopy (XPS) and inverse photoemission spectroscopy (IPS), which probe the entire occupied and unoccupied DOS, respectively. Unfortunately both of these techniques are extremely surface sensitive and require single crystal samples. In contrast, the X-rays used in the oxygen XES and XAS measurements have an attenuation length of on the order of a hundred nanometers\cite{henke93}, making them much more bulk sensitive. Since the oxygen states play the major role in the valence and conduction bands near the Fermi level, we use the oxygen XAS and XES spectra to estimate the bandgaps of these materials. 

Soft X-ray spectroscopy may be performed on almost any material which can be positioned in the X-ray beam, therefore demonstrating that XES and XAS are suitable techniques for estimating bandgaps makes this parameter much easier to determine. The bandgap is arguably one of the most important parameters in such materials, and the ability to tune the bandgap is desirable in order to tailor materials for use in specific devices. Ultimately, understanding the differences in the bandgap among these related materials may give insight into mechanisms for tuning the bandgap in other materials.

\section{Experimental and Calculation Details}

The oxygen XES measurements were performed at Beamline 8.0.1 of the Advanced Light Source (ALS) at Lawrence Berkeley National Laboratory (LBNL) \cite{ji95}. The endstation uses a Rowland circle geometry X-ray spectrometer with spherical gratings and an area sensitive multichannel detector. The oxygen \textit{K}$_\alpha$ XES were excited near the oxygen \textit{1s} ionization threshold, at 540.8 eV, to suppress the high-energy satellite structure. The spectrometer resolving power (E/$\Delta$E) for emission measurements was about 800.

The oxygen XAS measurements were performed at the Spherical Grating Monochromator (SGM) beamline of the Canadian Light Source (CLS) at the University of Saskatchewan \cite{regier07}. The absorption measurements were acquired in total fluorescence mode (TFY) using a channelplate fluorescence detector. The monochromator resolving power (E/$\Delta$E) for absorption measurements was about 2000. TFY provides more bulk sensitivity than the other common technique for soft X-ray absorption measurements, total electron yield (TEY). All spectra were normalized to the incident photon current using a highly transparent gold mesh in front of the sample to measure the intensity fluctuations in the photon beam.

To estimate the Fermi level and the bandgap of these materials in a consistent manner we make use of peaks in the second derivative of the absorption and emission spectra. Using local maxima in the second derivative has been shown to be a good tool for estimating the band gap in oxides \cite{kurmaev08}.

The metal oxides measured were all commercially available powders (Alfa Aesar, purity 99\%). These powders were mounted on clean indium foil and measured without further processing.

Our calculations were performed using the WIEN2k code \cite{blaha01}, a full-potential linear augmented plane-wave plus local-orbital (FP-LAPW+lo) method employing scalar-relativistic core orbital corrections. We used the generalized gradient approximation (GGA) of Perdew-Burke-Ernzerhof \cite{perdew96} for the exchange-correlation functional of the alkaline earth oxides. The post-transition metal oxides have \textit{3d}-orbitals, which are not properly represented with GGA functionals. To correct for this, we used the local spin density approximation Fock-$\alpha$ (LSDA-Fock-$\alpha$) \cite{moreira02} exchange-correlation functional for the post-transition metal oxides. The $\alpha$ parameter is a free variable, which may be chosen to reproduce the experimental results, but to keep these calculations as \textit{ab initio} as possible we chose a typical value of $\alpha = 0.35$ \cite{tran06} for all post-transition metal oxides. The Brillouin zone integrations were performed with a 1000 \textit{k}-point grid and \textit{R}$_{\mathit{MT}}^{\mathit{min}}$\textit{K}$_{\mathit{max}}$ = 7 (the product of the smallest of the atomic sphere radii \textit{R}$_{\mathit{MT}}$ and the plane wave cutoff parameter \textit{K}$_{\mathit{max}}$) was used for the expansion of the basis set. The space groups, lattice parameters, Brillouin zone \textit{k}-point axes, and atomic sphere radii used for all compounds is given in Table \ref{tbl:calc_details}. The atomic sphere radii were chosen such that the spheres were nearly touching.

\begin{table}
\begin{tabular}{cccccccccc}
 & Space & & & & Metal & Oxygen & \textit{k}-grid \\
Compound & Group & \textit{a} [\AA]& \textit{b} [\AA]& \textit{c} [\AA]& Site & Site & $k_a \times k_b \times k_c$ & \textit{R}$_{MT}^{metal}$ & \textit{R}$_{MT}^{O}$\\
\hline
BeO \cite{xu93}& P6$_3$mc & 2.698 & 2.698 & 4.380 & 2(b) & 2(b) & 12 $\times$ 12 $\times$ 6 & 1.53 & 1.53\\
MgO \cite{hazen76}& Fm$\bar{3}$m & 4.211 & 4.211 & 4.211 & 4(a) &4(b) & 10 $\times$ 10 $\times$ 10 & 1.96 & 1.96  \\
CaO \cite{fiquet99} & Fm$\bar{3}$m & 4.815 & 4.815 & 4.815 & 4(a) &4(b) & 10 $\times$ 10 $\times$ 10 & 2.24 & 2.24 \\
SrO \cite{wyckoff63} & Fm$\bar{3}$m & 5.160 & 5.160 & 5.160 & 4(a) &4(b) & 10 $\times$ 10 $\times$ 10 & 2.40 & 2.40 \\
BaO \cite{wyckoff63}& Fm$\bar{3}$m  & 5.523 & 5.523 & 5.523 & 4(a) &4(b) & 10 $\times$ 10 $\times$ 10 & 2.50 & 2.50\\
ZnO \cite{kinhara85}& P6$_3$mc & 3.2494 & 3.2494 & 5.2038 & 2(b) & 2(b) & 12 $\times$ 12 $\times$ 6 &1.77 & 1.57\\
CdO \cite{zhang99}& Fm$\bar{3}$m & 4.211 & 4.211 & 4.211 & 4(a) &4(b) & 10 $\times$ 10 $\times$ 10 & 1.96 & 1.96 \\
HgO \cite{aurivillius64} & Pnma & 6.6129 & 5.5208 & 3.5219 & 4(c) & 4(c) & 7 $\times$ 9 $\times$ 14 & 2.01 & 1.78
\end{tabular}
\caption{\label{tbl:calc_details}Crystal structure parameters used in theoretical calculations. All are based on experimental data. The space groups for BeO, ZnO, and HgO have an additional degree of freedom in the atomic sites. For BeO, these are z$_\mathrm{Be} = 0$ and z$_\mathrm{O} = 0.378$ \cite{xu93}. For ZnO these are z$_\mathrm{Zn} = 0$ and z$_\mathrm{O} = 0.3821$ \cite{kinhara85}. For HgO these are x$_\mathrm{Hg} = 0.1136$, z$_\mathrm{Hg} = 0.2456$, x$_\mathrm{O} = 0.3592$, and z$_\mathrm{O} = 0.5955$ \cite{aurivillius64}.}
\end{table}

The X-ray spectra for the oxygen \textit{K}-edge were calculated from the electronic structure using the ``XSPEC'' program included in the WIEN2k code. This utility estimates the X-ray spectra by multiplying the matrix of allowed dipole transitions with a radial transition probability and the appropriate partial density of states. The formalism for this method is outlined in Reference \onlinecite{schwarz79}. The calculated spectra were broadened by a Voigt function using the instrumental resolving power for the width of the Gaussian component and a variable Lorentzian width ranging from 0 eV at the edge of the conduction band to about 1 eV at the end of the spectrum following a quadratic curve, as outlined in Reference \onlinecite{goodings69}. The absorption spectra were calculated using a $2a \times 2b \times 2c$ super-cell with a single oxygen \textit{1s} electron removed and a background charge of e$^{-}$ added to account for the effect of the core-hole created in the photoexcitation process. Note that the calculated XAS do not include the photoionization of the excited atom, which manifests roughly as a step function in the measured XAS. This is why the higher energy states in the measured XAS have uniformly greater amplitudes (and there is a greater background signal) than the calculated XAS.

In the discussion that follows, the measured XES and XAS are displayed on the same energy scale, and the Fermi level is chosen as the local maxima in the second derivative on the high energy side of the XES. The calculated DOS, XES, and XAS spectra have then been aligned with this point, to facilitate a direct comparison between calculations and measurements.

\section{Results and Discussion}
In general we have obtained good qualitative agreement between measured and calculated X-ray spectra for these materials. As expected, the valence DOS for the materials studied herein is almost completely composed of oxygen \textit{2p}-states. Both the measured and calculated XAS of these materials share the same bulk features as the conduction DOS with no core-hole; the effect of the core-hole seems to distort the spectra by increasing the number of available states near the low-energy end of the conduction band rather than shift the band itself to lower energies; the latter effect is common in the \textit{L}$_\mathit{2,3}$ spectra of transition metals \cite{mauchamp09}. 

\newcounter{subfig}
\renewcommand{\thefigure}{\arabic{figure}\alph{subfig}}
\setcounter{subfig}{1}
\begin{figure}
\includegraphics[width=3in]{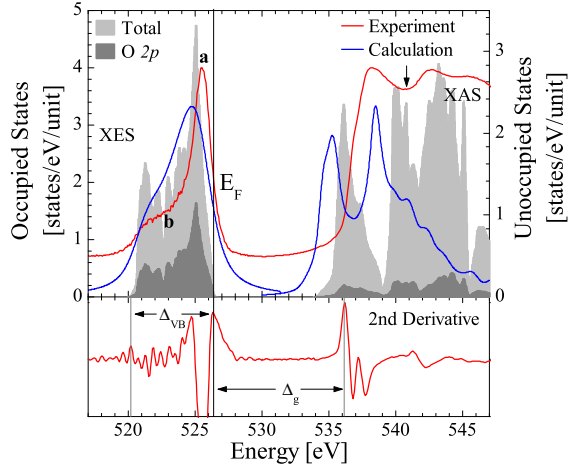}
\caption{\label{fig:beo}Measured oxygen \textit{K}-edge soft X-ray spectra, calculated X-ray spectra, and calculated density of states for BeO. The XES excitation energy, at 540.8 eV, is marked by an arrow at the corresponding location in the XAS. The Fermi level (E$_\mathrm{F}$) is denoted by a vertical line. The second derivative of the measured X-ray spectra is plotted at the bottom, with the bandgap $\Delta_\mathrm{g}$ and the valence bad width $\Delta_\mathrm{VB}$ labeled. Note that the scale for the DOS varies; for the total DOS it is in [states/eV/unit cell] and for oxygen it is in [states/eV/atom]. Both the occupied and unoccupied ground state DOS is displayed here, only the calculated XAS makes use of the core-hole calculation. (Colour in online version.)}
\end{figure}

The measurements and calculated spectra for BeO are shown in Figure \ref{fig:beo}. The valence band matches the measured XES quite well, save that the post-edge features (near the XES shoulder at \textbf{b}) are populated slightly higher in the calculated DOS than they appear relative to the main spectral line (the peak at \textbf{a}) in the measured XES. Unlike the other alkaline earth oxides, the core-hole causes the calculated XAS to be shifted by about 1 eV to lower energies. This is easily explained: the conduction band states in all the alkaline earth oxides are far less localized than the valence states, and there is considerable hybridization with the states from the alkaline metal. Beryllium is the only alkaline earth that is lighter than oxygen so the effective increased nuclear charge from an oxygen core-hole has a much greater effect on the neighbouring beryllium states, and thus the local conduction band in general, than an oxygen core-hole in a system with a more massive cation.

\addtocounter{subfig}{1}
\addtocounter{figure}{-1}
\begin{figure}
\includegraphics[width=3in]{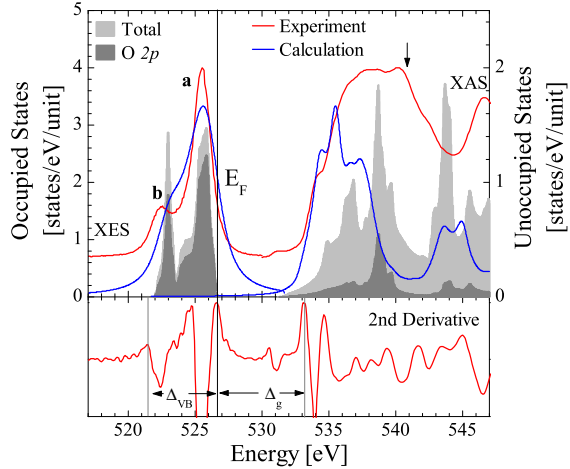}
\caption{\label{fig:mgo}Measured oxygen \textit{K}-edge soft X-ray spectra, calculated X-ray spectra, and calculated density of states for MgO. This Figure has the same format as Figure \ref{fig:beo}. (Colour in online version.)}
\end{figure}

The measurements and calculated spectra for MgO are shown in Figure \ref{fig:mgo}. Here again the XES shows two distinct bands (\textbf{a} and \textbf{b}), the secondary band \textbf{b} is slightly more prominent and better separated from the main band \textbf{a} than the bands in BeO. Compared to the calculated DOS, both the measured and calculated XAS indicate a shifting of the spectral weight to states closer to the edge of the conduction band, but the edge of the calculated XAS remains at the edge of the conduction band without the core-hole. This type of core-hole behaviour has also been reported for silicon \textit{K}-edge XAS \cite{weijs90}. We also point out that the agreement between the calculated and measured XAS for MgO (and BeO) is rather poor, in particular it appears that too much emphasis is placed on near-edge conduction states. For example, the sharp feature at $\sim$536 eV in the calculated XAS is only a very minor lump in the broad onset of the measured XAS (note that the corresponding measured feature occurs at $\sim$538 eV, since the onset of the measured XAS is $\sim$2 eV higher than that of the calculated XAS). This suggests that either the O \textit{1s} core-hole is more screened than predicted by the calculation, or that our supercell was not large enough to sufficiently separate the core-holes. In the former case the situation may be remedied by introducing a partial core hole, in the latter simply expanding the supercell will suffice. Previous reports on calculating the XAS of MgO suggest that a full core-hole overestimates the attractive potential of the excited atom \cite{mizoguchi00}. We do not attempt to improve the XAS calculation for MgO (or BeO) because we wish to use a consistent method for all calculated XAS herein. While the full core-hole approach does not work well with the lighter oxides, it gets progressively better as the cations increase in mass. Finally, it should also be noted that the pre-edge XAS feature at $\sim$534 eV (in the measured XAS) is due to a surface effect and is not related to the bulk DOS \cite{pascual02}.

\addtocounter{subfig}{1}
\addtocounter{figure}{-1}
\begin{figure}
\includegraphics[width=3in]{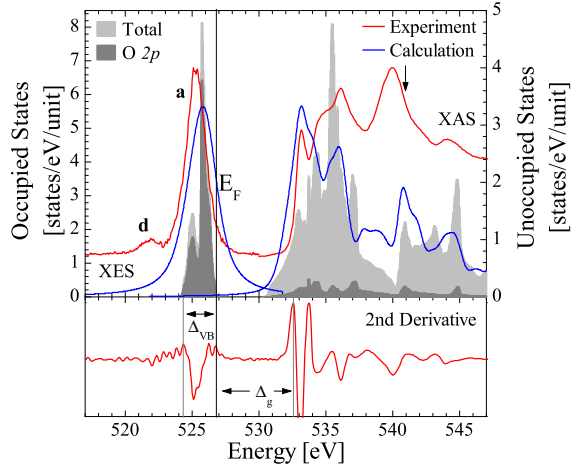}
\caption{\label{fig:cao}Measured oxygen \textit{K}-edge soft X-ray spectra, calculated X-ray spectra, and calculated density of states for CaO. This Figure has the same format as Figure \ref{fig:beo}. (Colour in online version.)}
\end{figure}

The measurements and calculated spectra for CaO are shown in Figure \ref{fig:cao}. The sample was prepared at ambient conditionals, and the secondary band at \textbf{d} is due to a small amount of CaCO$_3$ created via interactions with atmospheric CO$_2$. Since CaCO$_3$ has a rather intense secondary band compared to the primary one, the relative weakness of the feature at \textbf{d} indicates that the contamination is minor. It has also been noted that the secondary emission from the CO$_3^{-2}$ complex is enhanced with resonant excitation \cite{butorin94}. As previously noted we used resonant excitation to suppress the high energy satellite structures that occur in these materials; the low intensity of the CO$_3^{-2}$ feature even with resonant excitation implies a very minor CaCO$_3$ contamination.

\addtocounter{subfig}{1}
\addtocounter{figure}{-1}
\begin{figure}
\includegraphics[width=3in]{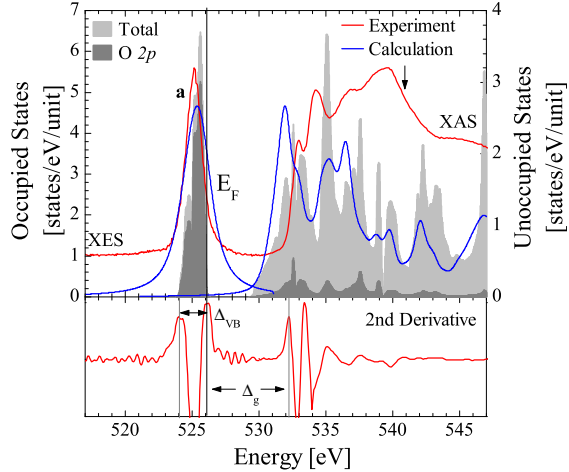}
\caption{\label{fig:sro}Measured oxygen \textit{K}-edge soft X-ray spectra, calculated X-ray spectra, and calculated density of states for SrO. This Figure has the same format as Figure \ref{fig:beo}. (Colour in online version.)}
\end{figure}

The measurements and calculated spectra for SrO are shown in Figure \ref{fig:sro}. There is a bit of an anomaly compared to the other compounds in this series: there is no secondary emission band at a lower energy than the primary band \textbf{a}. The most likely culprit is a small contamination of SrCO$_3$, for the same reasons as discussed above. The primary band is, however, in excellent agreement with the calculated XES and valence DOS. As previously discussed, the calculated XAS has greater spectral weight near the edge of the conduction band than the DOS without the core-hole, and following the trend in these materials the distortions from the core-hole are even weaker than in CaO, due to the increased stability in the conduction states provided by the progressively more massive cations.

\addtocounter{subfig}{1}
\addtocounter{figure}{-1}
\begin{figure}
\includegraphics[width=3in]{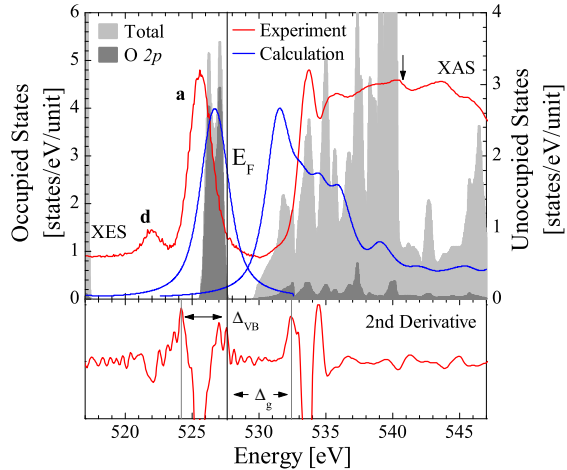}
\caption{\label{fig:bao}Measured oxygen \textit{K}-edge soft X-ray spectra, calculated X-ray spectra, and calculated density of states for BaO. This figure has the same format as Figure \ref{fig:beo}. (Colour in online version.)}
\end{figure}

The measurements and calculated spectra for BaO are shown in Figure \ref{fig:bao}. As in CaO, the main XES feature (band \textbf{a}) matches the calculated spectra and the secondary emission band \textbf{d} in the measured XES is due the CO$_3^{-2}$ complex formed by surface carbonation (BaCO$_3$). In addition to the previously discussed distortions present in the XAS of these materials, it is worth pointing out that the near-edge features in the BaO and SrO XAS are much sharper than those in BeO and MgO. This further suggests that the larger cations provide a stabilizing effect on the conduction band, in this case by increasing the core-hole life-time and subsequently improving the energy resolution.

In all alkaline earth oxides, the occupied metal \textit{ns}- and \textit{(n-1)d}-states have the same band width as the oxygen \textit{2p}-states (see Figure \ref{fig:beo} etc.). Further, the main spectral weight of the \textit{ns}-states hybridize with the low energy oxygen \textit{2p} band, while the metal \textit{(n-1)d}-states are weighted closer to the Fermi level.

\setcounter{subfig}{1}
\begin{figure}
\includegraphics[width=3in]{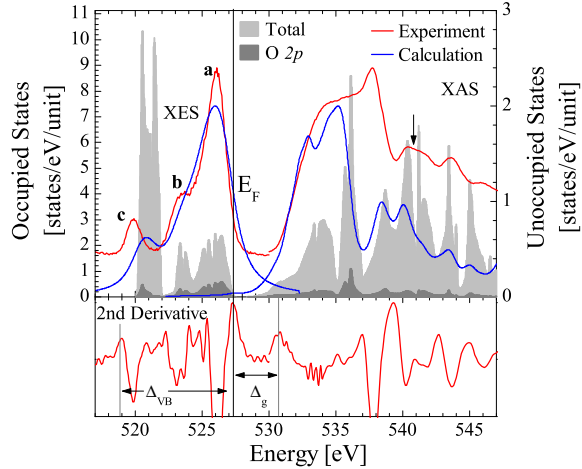}
\caption{\label{fig:zno}Measured oxygen \textit{K}-edge soft X-ray spectra, calculated X-ray spectra, and calculated density of states for ZnO. This Figure has the same format as Figure \ref{fig:beo}. (Colour in online version.)}
\end{figure}

The measurements and calculated spectra for ZnO are shown in Figure \ref{fig:zno}. Like the alkaline earth oxides, the valence band near the Fermi level in post-transition metal oxides is dominated by oxygen \textit{2p}-states, as seen in the shape of the calculated DOS and the two XES bands \textbf{a} and \textbf{b} near the Fermi level. In the post-transition metal oxides there is an additional band at \textbf{c}, formed by hybridized oxygen and metal \textit{nd}-states (\textit{n} = 3, 4, 5). As in BeO, the calculated DOS seem to over-estimate the population of the XES band at \textbf{b} - in the calculated XES this band is only part of a broad shoulder smoothly merging with the main band at \textbf{a}.

\addtocounter{subfig}{1}
\addtocounter{figure}{-1}
\begin{figure}
\includegraphics[width=3in]{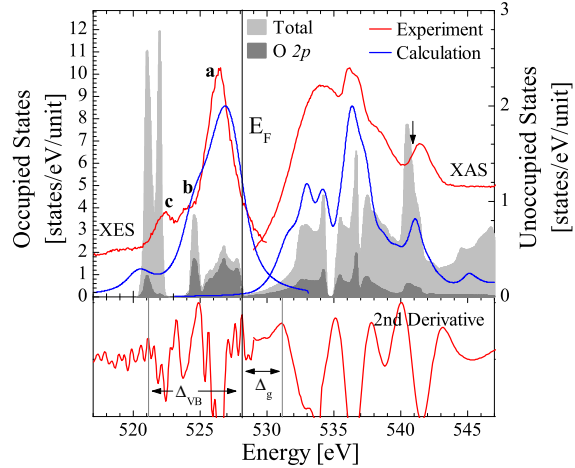}
\caption{\label{fig:cdo}Measured oxygen \textit{K}-edge soft X-ray spectra, calculated X-ray spectra, and calculated density of states for CdO. This Figure has the same format as Figure \ref{fig:beo}. (Colour in online version.)}
\end{figure}

The measurements and calculated spectra for CdO are shown in Figure \ref{fig:cdo}. While the calculated DOS predicts the hybridized oxygen \textit{2p}- and cadmium \textit{4d}-states in the same location as the hybridized oxygen \textit{2p}- and zinc \textit{3d}states in ZnO, here the XES band \textbf{c} is at a slightly higher energy than the calculated DOS, in contrast to the ZnO XES band \textbf{c} which is at a slightly lower energy than the calculated DOS. The agreement between measurement and calculation is still quite good, however, considering that these calculations used a Fock parameter $\alpha = 0.35$ simply because this result was shown to produce good results for a few transition metal oxides \cite{tran06}.

\addtocounter{subfig}{1}
\addtocounter{figure}{-1}
\begin{figure}
\includegraphics[width=3in]{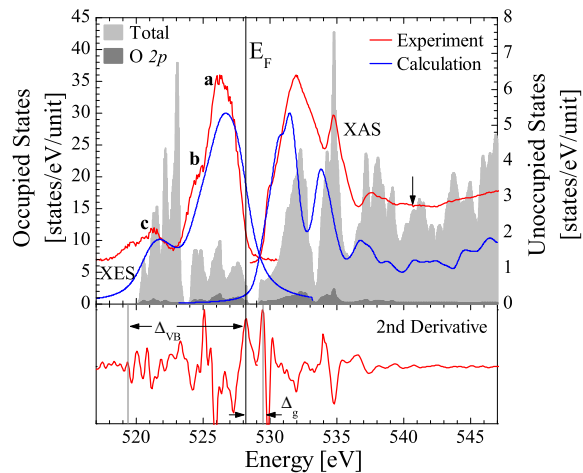}
\caption{\label{fig:hgo}Measured oxygen \textit{K}-edge soft X-ray spectra, calculated X-ray spectra, and calculated density of states for HgO. This Figure has the same format as Figure \ref{fig:beo}. (Colour in online version.)}
\end{figure}

The measurements and calculated spectra for HgO are shown in Figure \ref{fig:hgo}. Here the hybridized oxygen \textit{2p}- and mercury \textit{5d}-states form a much broader band in the calculated DOS, and the emission band \textbf{c} is broader than that of ZnO and CdO. The conduction DOS and the core-hole distorted XAS follow a similar trend to that of the alkaline earth oxides: as the cation increases in mass the core-hole has a decreasing effect on the local conduction band probed by XAS measurements. Indeed the agreement between the measured XAS and the calculated DOS is quite good for CdO, and excellent for HgO.

Like the alkaline earth oxides, the width of the occupied metal \textit{ns}- and \textit{(n-1)d}-states is the same as that of the oxygen \textit{2p}-states, (see Figure \ref{fig:zno} etc.). Unlike the alkaline earth oxides however, the low energy side of the band is dominated by hybridization between the oxygen \textit{2p}-states and the metal \textit{(n-1)d}-states, rather than the metal \textit{ns}-states.

Comparing the calculated and measured XES in these materials shows that choosing the local maximum in the second derivative of the XES is a reasonable guess at the Fermi level of these materials. In all materials except BeO, using the second derivative to estimate the Fermi level puts the measured XES (indexed by peak \textbf{a}) at a slightly lower energy than the calculated XES. BaO has the greatest discrepancy; the measured XES is 1.2 eV lower than the calculated XES. The other materials all have less than 1 eV difference between the measured and calculated XES, the next largest discrepancy is the 0.8 eV difference between the calculated and measured XES for BeO.

It is apparent that in these materials many of the calculated XAS are shifted to lower energies than the measured XAS; often even more so if the calculated XES is aligned to the measured XES. This reflects the well known fact that GGA methods usually underestimate the bandgap \cite{dufek94}. To further complicate the problem, there some disagreement in the literature on what the bandgaps could be. The results from our calculations, measurements, and literature survey for the bandgaps are recorded in Table \ref{tbl:gap}. We emphasize that our experimental results are mostly comparable with the experimental results reported in the literature. 

The presence of a core-hole in the XAS measurement process distorts the measured XAS from the true, unperturbed, conduction band. It is also possible that the core-hole can shift the conduction states local to the excited atom to lower energies, this often occurs in the \textit{2p} XAS of metals (see, for example Reference \onlinecite{kurmaev08}). If the latter effect is prominent in O \textit{1s} XAS, then obviously the XES and XAS alignment is not a useful probe for the bandgap. Fortunately, our calculations (with the exception of BeO, as previously mentioned) show that while the core-hole causes a greater density of low energy conduction band states, it does not change the onset energy of the conduction band nor substantially alter the energies of prominent conduction band features (although the relative intensities of these features may be changed substantially). This effect was investigated in detail by Cabaret \textit{et al.} in germanates\cite{cabaret07}.

Our XAS measurements are in good agreement with those available in the literature. Most studies are fairly dated, and there is a fair amount of discrepancy. See, for example, the summary in Reference \onlinecite{fronzoni05}. MgO, Ca, SrO, and BaO have previously been measured in TEY mode or by electron energy loss spectroscopy (EELS) \cite{lindner86,nakai87,weng89}. Both these techniques are highly surface sensitive, this accounts for the discrepancies between measurements. Since these studies are over 20 years old, given the progress in beamlines and synchrotron sources, and the fact that our measurements herein are much more bulk sensitive, we can regard our measurements as more accurate. ZnO XAS has been reported more recently \cite{guo02}, and is essentially the same as the spectrum we report here.

\begin{table}
\begin{tabular}{cccc}
Compound & $\Delta_\mathrm{g}$ [eV] & G$_\mathrm{c}$ [eV] & Literature Bandgaps [eV]\\
\hline
BeO & 9.9 & 8.4 & 6.4 --- 7.0 \cite{chang83}, 7.54 \cite{xu93}, 8.4 \cite{milman01}, 10.6$^\dagger$ \cite{emeline99}, $\>$10$^\dagger$ \cite{roessler69}\\
MgO & 6.4 & 4.8 & 4.8 \cite{kalpana95}, 7.4 \cite{bredow00}, 7.6 \cite{mathon01}, 7.7$^\dagger$ \cite{schonberger95}, 7.8$^\dagger$ \cite{roesler67}, 7.83$^\dagger$ \cite{whited73}, 8.7$^\dagger$ \cite{emeline99}\\
CaO & 5.7 & 3.7 & 7.09$^\dagger$ \cite{whited73}, 7.72 \cite{kotani94}, 7.73 \cite{pandey90}\\
SrO & 6.1 & 3.4 & 4.1 \cite{kalpana95}, 5.9$^\dagger$ \cite{rao79}\\
BaO & 4.8 & 2.0 & 1.75 \cite{junquera03}, 4.8$^\dagger$ \cite{strewlow73}, 5 \cite{mckee01}\\
\hline
ZnO & 3.4 & 1.3 & $\sim$1 \cite{azzaz08}, 3.3$^\dagger$ \cite{ozgur05, srikant98}, 3.37$^\dagger$ \cite{klingshirn75}\\
CdO & 2.9 & 0.7 & 0.55$^\dagger$ \cite{kohler72}, 2.06$^\dagger$ \cite{carballeda00}, 2.22$^\dagger$ \cite{ueda98},  2.46$^\dagger$ \cite{subramanyam97}\\
HgO & 1.2 & 1.0 & 1.9$^\dagger$ \cite{glans05}
\end{tabular}
\caption{\label{tbl:gap}Measured, calculated, and previously reported bandgaps for all materials studied herein. The measured bandgaps are denoted by $\Delta_\mathrm{g}$ and the calculated bandgaps are denoted by G$_\mathrm{c}$; the symbols are explicitly different to prevent confusion. The literature bandgaps labeled with $^\dagger$ are based on experiment, the others on calculations.}
\end{table}

While there is obviously a discrepancy between our estimates of the bandgaps from measured X-ray spectra and DFT calculations, it is important to point out that there is a correlation between them; the calculated bandgaps are all roughly 1.5 eV smaller than the bandgap estimates from X-ray spectra. Figure \ref{fig:gap_compare}, left hand side, shows the calculated gap G$_\mathrm{c}$ plotted with respect to the measured gap $\Delta_\mathrm{g}$. If the calculations could reproduce the measured spectra perfectly, all points would fall along the line G$_\mathrm{c} = \Delta_\mathrm{g}$ (denoted by the gray line). As it stands, the relationship is linear (the black dotted curve is the best-fit line), and it is encouraging that the slope is quite close to unity, especially since both methods of calculating the bandgap are completely independent of each other. 

The calculations are much more accurate at predicting the occupied valence band, and as shown in Figure \ref{fig:gap_compare}, right hand side, the widths of the occupied DOS W$_\mathrm{VB}$ and the measured XES $\Delta_\mathrm{VB}$ (as denoted by maxima in the second derivative) are basically the same. It has been suggested (in the case of MgO) that calculations based on the GW approximation can improve the match between W$_\mathrm{VB}$ and $\Delta_\mathrm{VB}$ \cite{shirley98}, but the improvement over conventional GGA calculations is minor. We should add that other methods of calculating electronic structure, such as the GW approximation \cite{cappellini00} and correlated Hartree-Fock methods \cite{pandey91}, can achieve better agreement with experiment. However these calculations are more intensive than the DFT-GGA method and the best results require atom-specific pseudopotentials which are based on carefully selected ground state electron occupations \cite{shirley98}. Our experimental results show that a simple GGA calculation is accurate in reproducing the valence bands of alkaline earth and post-transition metal oxides, and that the error in the calculated bandgap is roughly a constant shift.

\renewcommand{\thefigure}{\arabic{figure}}
\begin{figure}
\includegraphics[width=3in]{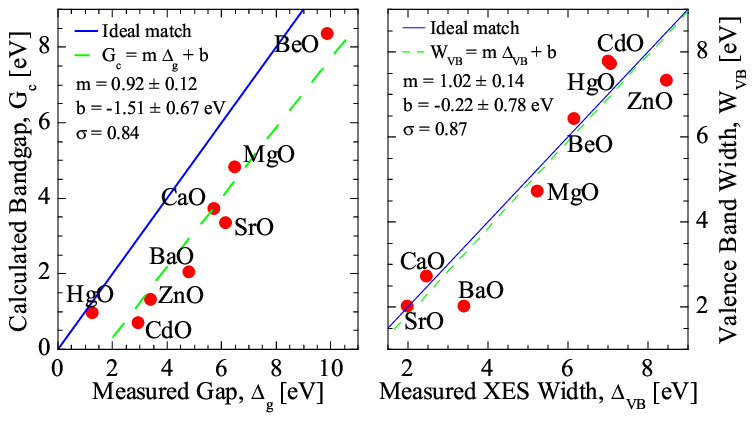}
\caption{\label{fig:gap_compare}The left-hand side shows the calculated bandgap G$_\mathrm{c}$ plotted with respect to the measured gap $\Delta_\mathrm{g}$. The right-hand side shows the width of the calculated valence band W$_\mathrm{VB}$ plotted with respect to the width of the measured XES $\Delta_\mathrm{VB}$. In both plots the gray line shows the ideal relationship if the calculations perfectly matched the measurement, and the dotted black line shows the best linear fit through the data. (Colour in online version.)}
\end{figure}

\renewcommand{\thefigure}{\arabic{figure}\alph{subfig}}
\setcounter{subfig}{1}
\begin{figure}
\includegraphics[width=3in]{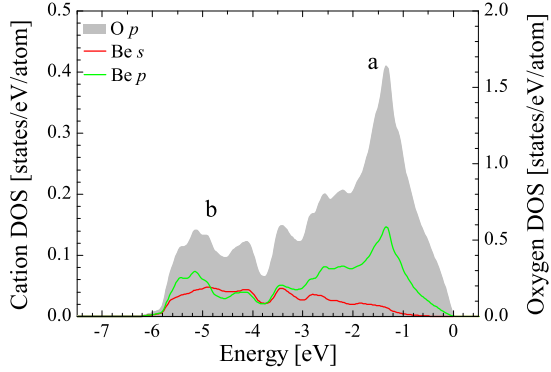}
\caption{\label{fig:bedos}Calculated oxygen \textit{2p} and beryllium \textit{2s} and \textit{2p} valence band DOS for BeO. Note the oxygen DOS is scaled differently than the metal DOS. (Colour in online version.)}
\end{figure}

\addtocounter{subfig}{1}
\addtocounter{figure}{-1}
\begin{figure}
\includegraphics[width=3in]{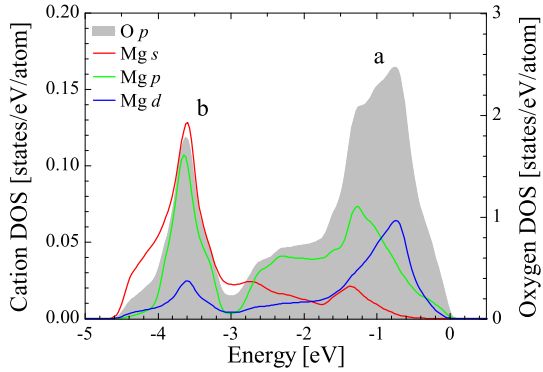}
\caption{\label{fig:mgdos}Calculated oxygen and magnesium DOS for MgO. This Figure has the same format as Figure \ref{fig:bedos}. (Colour in online version.)}
\end{figure}

The calculated electronic structure for BeO and MgO reveal that the primary valence band \textbf{a} is composed of oxygen \textit{2p} states hybridized primarily with metal \textit{2p} states (see Figures \ref{fig:bedos} and \ref{fig:mgdos}). Likewise the secondary valence band \textbf{b} is composed of oxygen \textit{2p} states hybridized primarily with metal \textit{2s,p} states. We can therefore classify the \textbf{a}-band as \textit{p}-like, and the \textbf{b}-band as \textit{sp}-like. Since band structure calculations of metallic Be \cite{chou83,haussermann01}, Mg \cite{haussermann01,gotsis02} and Zn \cite{haussermann01} show that the metal \textit{p} and \textit{s} bands are divided in the same way as those in BeO, MgO, and ZnO, this separation in metal bands is inherent to the cation, rather than a side effect of oxygen hybridization.

\addtocounter{subfig}{1}
\addtocounter{figure}{-1}
\begin{figure}
\includegraphics[width=3in]{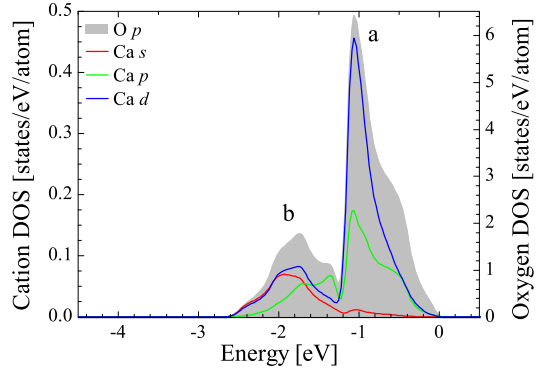}
\caption{\label{fig:cados}Calculated oxygen and calcium DOS for CaO. This Figure has the same format as Figure \ref{fig:bedos}. (Colour in online version.)}
\end{figure}

\addtocounter{subfig}{1}
\addtocounter{figure}{-1}
\begin{figure}
\includegraphics[width=3in]{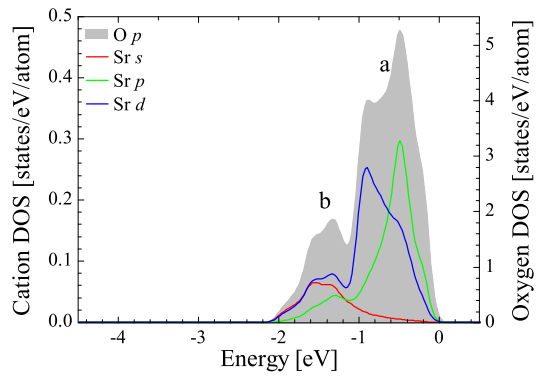}
\caption{\label{fig:srdos}Calculated oxygen and strontium DOS for SrO. This Figure has the same format as Figure \ref{fig:bedos}. (Colour in online version.)}
\end{figure}

\addtocounter{subfig}{1}
\addtocounter{figure}{-1}
\begin{figure}
\includegraphics[width=3in]{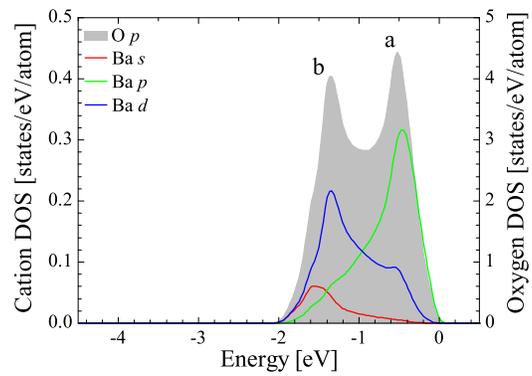}
\caption{\label{fig:bados}Calculated oxygen and barium DOS for BaO. This Figure has the same format as Figure \ref{fig:bedos}. (Colour in online version.)}
\end{figure}

The energy difference between the \textbf{a} and \textbf{b} valence bands is about 3.9 eV for BeO and 2.5 for MgO. In the remaining alkaline earth oxides this difference is 1.0 eV for CaO, 0.8 eV for SrO, and 0.8 eV for BaO (see Figures \ref{fig:cados}, \ref{fig:srdos}, and \ref{fig:bados}) and the secondary band \textbf{b} is not resolved in the measured XES (refer back to Figures \ref{fig:cao}, \ref{fig:sro}, \ref{fig:bao}). The general reduction in the energy difference is predicted by basic atomic theory: the energy of the \textit{ns} electrons is roughly $E \sim \left(Z - \sigma\right)^2 n^{-2}$ (where $Z$ is the atomic number and $\sigma$ is the screening parameter)\cite{blokhin57}. With increasing \textit{n}, the average energy of the \textit{s} states draws closer to the Fermi level (see Figure \ref{fig:aeo_states}). Since the Fermi level is the boundary for occupied states, decreasing energy of the \textit{s} states consequently causes the width of the states to become narrower, which decreases the range hybridization with oxygen and causes the decrease in the spacing between the \textbf{a} and \textbf{b} regions. Indeed, Figure \ref{fig:aeo_states} suggests that the average energy of the \textit{np} states follow the same trend as the \textit{ns} states; decreasing with increasing  \textit{n}. We should also point out that CaO, SrO, and BaO have much larger (\textit{n-1})\textit{d} contributions to the valence band, in fact for these materials we should refer to the \textbf{a} valence band as \textit{pd}-like and the \textit{b} valence band as \textit{spd}-like.

As shown in Figure \ref{fig:aeo_states}, the energy of the cation states in MgO increases in energy (i.e. approaches the Fermi level) according to orbital symmetry: \textit{3s}, \textit{3p}, then \textit{3d}. Since subsequent alkaline earths have \textit{(n-1)d}-states they are at lower energies than the cation \textit{np}-states. These atomic-like tendencies in cation states cause the bulk of the O \textit{2p}-states to be located at the same energy as the cation \textit{3d}-states in MgO, and at the same energy as the cation \textit{np}-states in CaO, SrO, and BaO. Note that in CaO the cation \textit{4p}-states and \textit{3d}-states occur at the same energy.

\renewcommand{\thefigure}{\arabic{figure}}
\begin{figure}
\includegraphics[width=2in]{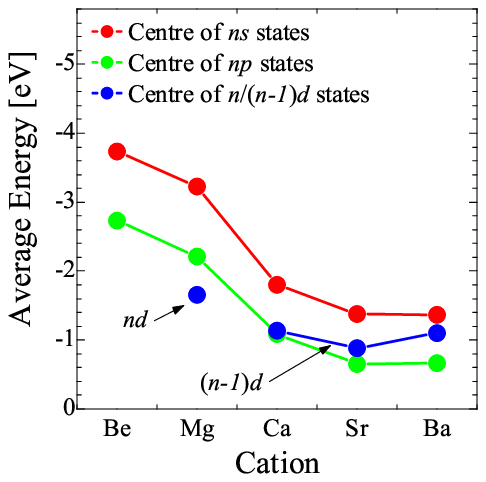}
\caption{\label{fig:aeo_states}The energy of the centre point (or ``centre of gravity'') of the cation valence \textit{ns}, \textit{np}, and \textit{nd}/(\textit{n-1})\textit{d} states for the alkaline earth oxides. (Colour in online version.)}
\end{figure}

\renewcommand{\thefigure}{\arabic{figure}\alph{subfig}}
\setcounter{subfig}{1}
\begin{figure}
\includegraphics[width=3in]{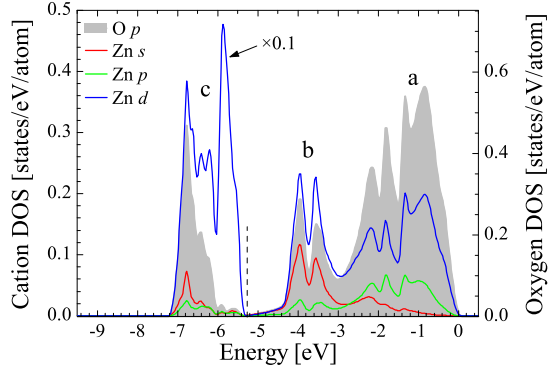}
\caption{\label{fig:zndos}Calculated oxygen \textit{2p} and zinc \textit{4s}, \textit{4p}, and \textit{3d} valence band DOS for ZnO. Note the oxygen DOS is scaled differently than the metal DOS. The \textit{3d} states at feature \textbf{c} have been scaled by a factor of 0.1, since otherwise the cation states near the Fermi level would be too small to see. (Colour in online version.)}
\end{figure}

\addtocounter{subfig}{1}
\addtocounter{figure}{-1}
\begin{figure}
\includegraphics[width=3in]{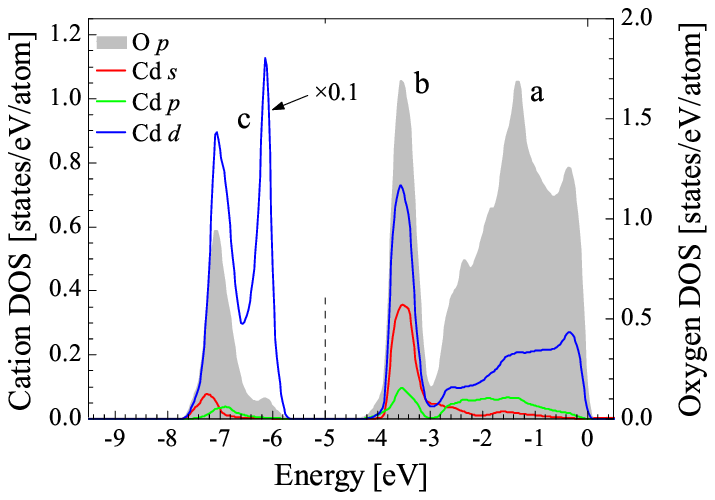}
\caption{\label{fig:cddos}Calculated oxygen and cadmium DOS for CdO. This Figure has the same format as Figure \ref{fig:zndos}. (Colour in online version.)}
\end{figure}

\addtocounter{subfig}{1}
\addtocounter{figure}{-1}
\begin{figure}
\includegraphics[width=3in]{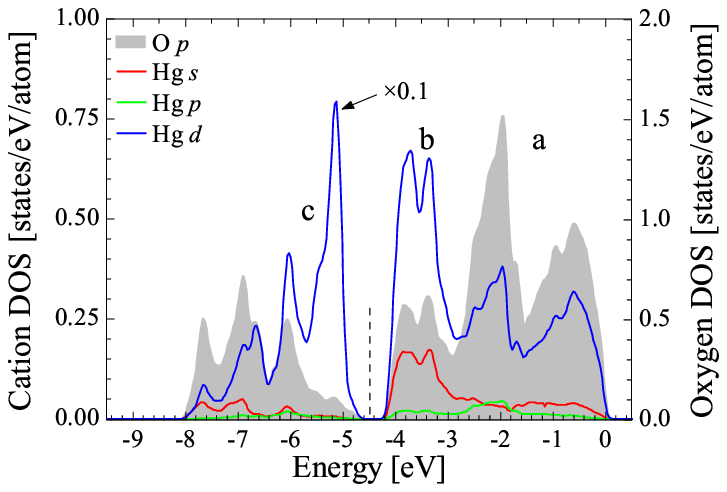}
\caption{\label{fig:hgdos}Calculated oxygen and mercury DOS for HgO. This Figure has the same format as Figure \ref{fig:zndos}. (Colour in online version.)}
\end{figure}

Both the calculated electronic structure and measured XES for the post-transition metal oxides show three valence bands (see Figures \ref{fig:zndos}, \ref{fig:cddos}, and \ref{fig:hgdos}). The additional valence band \textbf{c} is due to the (\textit{n-1})\textit{d} states of the cations. As expected, the valence band \textbf{c} is split due to the spin-orbit interaction within the cation (\textit{n-1})\textit{d} states, this splitting increases with increasing \textit{n}. Apart from the significant cation \textit{d} contribution, the valence band \textbf{b} is primarily \textit{ns}-like, like the alkaline earth oxides. Further, the distribution of \textit{np} states are shifted deeper in energy from ZnO to HgO and diminished in magnitude; indeed it is no longer appropriate to claim that the valence band \textbf{a} is \textit{pd}-like for HgO since the \textit{np} contribution is so small, now it is \textit{d}-like. In the region of band \textbf{b}, the \textit{d} DOS for the elements of the IIb group is significantly greater than those of the IIa group. We propose that this is due to \textit{nd}-(\textit{n-1})\textit{d} hybridization within the cation sphere.

\renewcommand{\thefigure}{\arabic{figure}\alph{subfig}}
\setcounter{subfig}{1}
\begin{figure}
\includegraphics[width=3in]{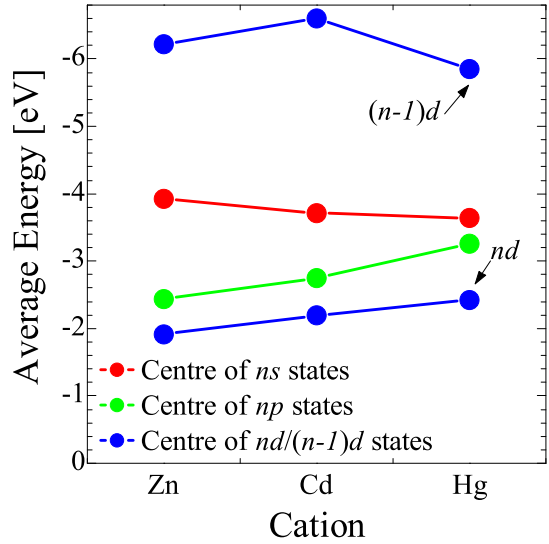}
\caption{\label{fig:ptmo_states}The energy of the centre point (or ``centre of gravity'') of the cation valence \textit{ns}, \textit{np}, and \textit{nd}/(\textit{n-1})\textit{d} states for the post-transition metal oxides. (Colour in online version.)}
\end{figure}

As shown in Figure \ref{fig:ptmo_states}, the cation partial DOS closest to the Fermi level is \textit{nd}-states, and as in the case of MgO, the energy of the cation states approach the Fermi level according to orbital symmetry: \textit{ns}, \textit{np}, then \textit{nd}. As before the main feature in the O \textit{2p} DOS is coincident in energy with the cation partial DOS closest to the Fermi level. The bulk of the O \textit{2p}-states coincide with the cation \textit{np}-states in BeO, SrO, and BaO, with the cation \textit{4p}- and (\textit{3})\textit{d}-states in CaO, and with the \textit{nd}-states in MgO, ZnO, CdO, and HgO. As a result of \textit{Me}-O hybridization the O \textit{2p}-states are mixed with all bonding \textit{s,p,d}-states of the cation in the aforementioned \textbf{a}, \textbf{b} and \textbf{c} subbands.


To summarize, we have measured the oxygen \textit{K}-edge emission and absorption X-ray spectra and calculated the DOS and X-ray spectra for the alkaline earth oxides \textit{Ae}O (\textit{Ae} = Be, Mg, Ca, Sr, Ba) and the post transition metal oxides \textit{Me}O (\textit{Me} = Zn, Cd, Hg). The bandgap estimates obtained from the XES and XAS follow the same trend as those obtained from calculations, save the measured bandgaps are on average 1.5 eV greater, which is expected since GGA is known to underestimate bandgaps. Both XES and calculations give essentially the same width for the valence band for all compounds. We have additionally found that the effect of the core-hole on XAS spectra becomes progressively weaker as the cation becomes more massive. Finally, by comparing the electronic states of the cation to the oxygen \textit{2p}-states we suggest that the oxygen \textit{K}-edge XES reflects (1) the distribution of the cation \textit{d}- and \textit{p}-states in band \textbf{a} with a sharp decrease in the role of \textit{p}-states in IIb group oxides relative to IIa group oxides and (2) the distribution of the cation \textit{d}- and \textit{s}-states in band \textbf{b}. The energy of the main O \textit{K}-edge XES feature is defined by the symmetry of the cation partial DOS closest to the Fermi level (either \textit{p}- or \textit{d}-states), and likewise the main maximum of the O \textit{2p} DOS is located at the energy of appropriate cation partial DOS.

\begin{acknowledgments}
We acknowledge support of the Natural Sciences and Engineering Research Council of Canada (NSERC), the Canada Research Chair program, the Research Council of the President of the Russian Federation (Grants NSH-1929.2008.2 and NSH-1941.2008.2), and the Russian Science Foundation for Basic Research (Project 08-02-00148). M. Abu-Samak would like to thank the International Atomic Energy Agency (IAEA) for their financial support.
\end{acknowledgments}

\bibliography{metal_oxide}

\end{document}